\documentclass[11pt]{article}

\usepackage[preprint]{acl}

\usepackage{times}
\usepackage{latexsym}
\usepackage{amsmath}
\usepackage{enumitem}

\usepackage[T1]{fontenc}

\usepackage[utf8]{inputenc}

\usepackage{microtype}

\usepackage{inconsolata}

\usepackage{graphicx}

\title{\textsc{Curator}: Scalable Fine-grained Evidence Selection for Retrieval-Augmented Generation using Decision Models}

\author{Shib Dasgupta \\
  Amazon Web Services \\
  \texttt{shibdg@amazon.com} \\\And
  Bandish Parikh \\
  Amazon Web Services \\
  \texttt{bandish@amazon.com} \\ \And
  Rashmi Gangadharaiah \\
  Amazon Web Services \\
  \texttt{rgangad@amazon.com} \\}

\begin{document}
\maketitle 
\begin{abstract}

Retrieval-augmented generation (RAG) provides entire retrieved passages to the LLM, even though only a handful of spans within them contain the evidence pertaining to the question. While precise selection of these spans can improve context utilization, identifying them in isolation is challenging. A span may appear irrelevant by itself, and deciding whether to retain it may require examining other retrieved information. One can evaluate each span's candidacy with the full retrieved context, but this is computationally inefficient as the same text must be processed repeatedly. 
Pretrained decision model can be used to efficient evaluate multiple spans simultaneously, allowing each decision to consider the surrounding spans as well.
However, our experiments show that their ability to identify relevant evidence deteriorates as the number of simultaneous decisions increases.

To address this, we introduce \textsc{CURATOR}, a fine-tuned decision model trained to make reliable evidence-selection decisions over large candidate sets. By training a decision model to select varying numbers of supporting spans from large candidate pools, CURATOR develops the capability to select evidence for RAG while retaining the computational efficiency of joint decision-making. We observe that CURATOR maintains a strong evidence discrimination capability as the number of candidates increases, generalizes across different tasks and document domains, and improves downstream RAG performance, particularly when the available context is limited.
\end{abstract}

\section{Introduction}
\label{sec:introduction}

Retrieval-augmented generation (RAG) enables large language models to
answer questions using information retrieved from external knowledge
sources \cite{3495724.3496517,3524938.3525306}. These retrieved external documents or passages helps LLM to rely less on parametric knowledge for knowledge-intensive tasks such as open-domain question answering \cite{izacard-grave-2021-leveraging}. However, these retrieved passages
often contain substantially more irrelevant information than is needed to answer the question at hand, e.g., background information, unrelated details on other topics, huge tables with only a few important cells, etc. Thus, providing the entire passage increases the number of tokens processed by the answering LLM and may make relevant evidence harder to use
\cite{liu-etal-2024-lost,jiang-etal-2024-longllmlingua}. This problem is amplified in agentic workflows, where redundant and unnecessary information consumes context that could otherwise accommodate future tool outputs and intermediate reasoning \cite{packer2024memgptllmsoperatingsystems,pmlr-v306-sun26x}.
This inefficiency motivates us to study how to select evidence at a finer granularity under a limited context budget. Rather than forwarding entire retrieved passages, we aim to retain only the sentences or spans that are useful for answering the question.
\cite{ICLR2024_bda88ed2}.

\begin{figure*}[t]
\centering
\includegraphics[width=\textwidth]
{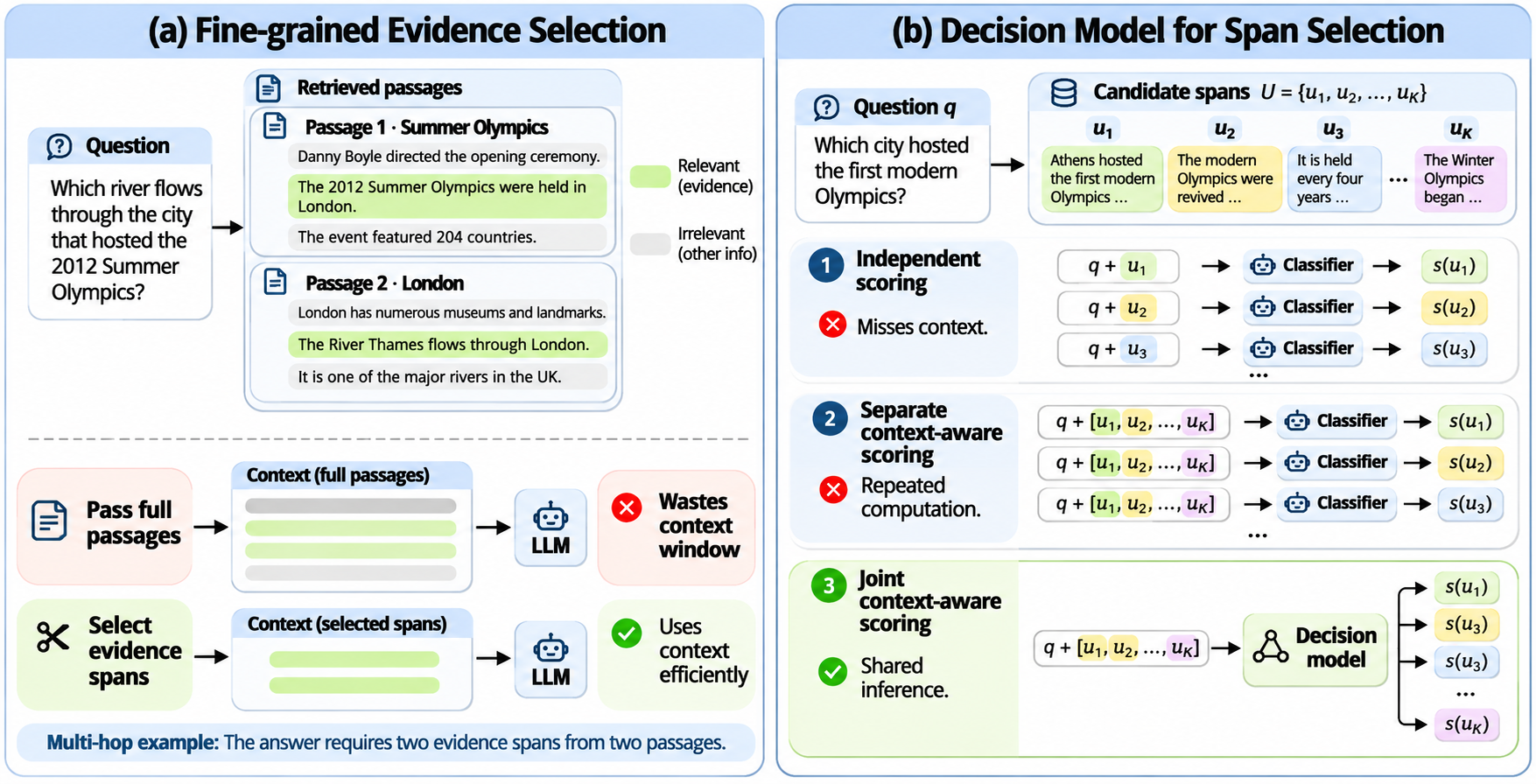}
\caption{
\textbf{Motivation for fine-grained evidence selection and joint scoring.}
(a) Retrieved passages contain relevant evidence alongside unnecessary information. Selecting useful spans reduces the input context passed to the answering model, leaving more room for other information within a limited context budget. (b) Three strategies for scoring candidate spans.
Independent scoring considers each span separately, without taking other candidates into account. Separate context-aware scoring accounts for all the spans
but requires repeatedly processing the shared context. Joint context-aware scoring evaluates multiple
candidates through shared inference, avoiding redundant computation over the full context.
}
\label{fig:curator_overview}
\end{figure*}

While fine-grained evidence selection can significantly reduce
input tokens to the answering model, the selection process
itself can be inefficient and introduce complexity if not
done correctly. A collection of retrieved passages may
contain hundreds of candidate spans. A simple approach is
to evaluate each candidate independently using the question
and the candidate span. However, a span may appear irrelevant when considered in isolation but become relevant when combined with information found elsewhere in the retrieved context. 

Figure~\ref{fig:curator_overview}(a) illustrates this
challenge through a two-hop question: one span establishes that London hosted the 2012 Summer Olympics, while another
establishes that the River Thames flows through London. Collectively, these spans provide the right evidence needed to
identify the river flowing through the city in question.
Neither span alone provides the complete answer. Alternatively, if we provide the entire retrieved context while judging each span then a model would theoretically be able to better judge the candidacy of each span jointly with other evidences. However, this will be computationally very inefficient as we would need to process same contextual information repetitively while judging each candidate span. 

Joint decision models offer a promising alternative. These models score predefined choices rather than generating a separate textual judgment for each candidate \cite{li2026llmasjevllmsjevstyledecision,geuter2026releasinginnerjevsimple}. Recent multi-readout decision models can produce many decisions through shared inference \cite{perplexity2026deciderflash}. This makes it possible to evaluate multiple evidence units together while allowing the information to be shared between candidates. Figure~\ref{fig:curator_overview}(b) illustrates three
span-scoring strategies for identifying such evidence,
motivating the use of joint decisions.

However, the ability to produce many decisions efficiently
does not necessarily guarantee their reliability. In our experiment, 
we show that pre-trained joint decision models
become less effective at distinguishing supporting evidence
from distractor spans as the number of candidate decisions
increases. Figure~\ref{fig:joint_scalability} illustrates this degradation on fact verification and multi-answer question answering tasks. This limitation is particularly important for fine-grained evidence selection for RAG, where a single retrieval step may yield many passages containing hundreds of candidate spans. That is why we investigate a core challenge in this paper: how can we preserve the computational advantages of joint inference while ensuring
that the decisions for selecting evidence remain reliable as the number of candidate spans increases?

To address this challenge, we introduce \textsc{CURATOR},
a framework for scalable joint evidence selection.
\textsc{CURATOR} adapts a pretrained multi-readout decision
model to identify supporting evidence spans across variable-sized
candidate sets. We construct supervised evidence-selection
examples from QAMPARI \cite{amouyal-etal-2023-qampari},
HotpotQA \cite{yang-etal-2018-hotpotqa}, and
2WikiMultiHopQA \cite{ho-etal-2020-constructing}, varying both the
number of candidates and the proportion of supporting
evidence present in those sets. This trains the model
to make many simultaneous decisions across different
evidence densities, including settings where only a few
candidates among many are relevant. We use class-balanced
joint evidence supervision to account for this imbalance.
To mitigate the loss of general decision capabilities,
we incorporate replay and distillation on general tasks
spanning binary classification, multiple-choice selection,
and scoring. At inference time, \textsc{CURATOR} jointly
scores every candidate evidence unit with the shared context. 
We then select the highest-scoring spans till the available token budget and leave out the rest. 

Through our experiments, we demonstrate the efficacy of
\textsc{CURATOR} in evidence span selection across different
tasks and domains. On QAMPARI, which is part of our training
data, \textsc{CURATOR} improves evidence recall by an average
of 25.8\% across different context budgets compared to
the pretrained joint decision model. The improvements
extend to unseen tasks and domains as well. For example,
we observe consistent improvements in evidence recall
on financial documents and legal contracts. Furthermore,
better evidence selection by \textsc{CURATOR} also leads
to improved downstream RAG performance on multi-hop
and financial question answering. The gains are
particularly pronounced when the available context
is limited. Our scalability experiments show that
\textsc{CURATOR} maintains strong evidence discrimination
as the number of simultaneous decisions increases,
unlike the pretrained model. These improvements are
achieved without any degradation
in the general decision-making capabilities of the
pretrained decision model. Our contributions are threefold:

\begin{itemize}[leftmargin=*, nosep]
\item \textbf{Efficient, context-aware evidence selection.}

We use joint decision models to select relevant evidence spans under a token budget, allowing multiple candidates to be evaluated together without repeatedly processing the full context.

\item \textbf{Scalable joint evidence selection decisions.}

We identify a degradation in evidence selection performance
as the number of candidate spans increases. As a remedy,
we introduce \textsc{CURATOR}, a framework that uses
variable-scale joint evidence supervision to improve
decision reliability for sparse evidence selection.

\item \textbf{Evaluation of quality, scalability,
and generalization.}

We demonstrate improved fine-grained evidence selection capability
with strong joint-decision scalability as the number of candidates increases. At a 128-token context budget,
CURATOR improves evidence recall by 17.0\% on average
across unseen dataset and domains, and downstream
RAG answer quality by 17.4\% on average across datasets relative to the strongest baselines.

\end{itemize}
\section{Problem Formulation and Motivation}
\label{sec:problem}
Retrieval-augmented generation (RAG) provides an answering model with retrieved passages, although only a fraction of the content may be needed to answer the question. Under a limited context budget, the task is not to simply rank passages but to decide \emph{which pieces of evidence to present} to the generator.
\subsection{Token-Budgeted Evidence Selection}
\label{sec:evidence_selection}

Given a question $q$ and retrieved passages $\mathcal{D}$,
let $U=(u_1,\ldots,u_K)$ denote the candidate evidence units
obtained from those passages. Here, a unit can be a sentence, multiple sentences, or a cell present in a table. Each unit $u_i$ has
an associated token cost $c_i=\operatorname{tokens}(u_i)$ and an
evidence-membership label $y_i\in\{0,1\}$.
Token-budgeted evidence selection aims to retain as much supporting evidence as possible within an allocated budget $B$. Formally, the selected subset of
indices $I_B\subseteq\{1,\ldots,K\}$ must satisfy
\begin{equation}
    \sum_{i\in I_B}c_i\leq B,
    \label{eq:evidence_budget}
\end{equation}

To achieve this, the selector model is trained to assign higher scores $r_i$ to supporting evidence than to irrelevant candidate units. At inference time, candidates are ranked by these scores and greedily admitted while adhering to the token budget. We keep this selection procedure fixed across methods and investigate how the scores are obtained.

\subsection{The Accuracy--Efficiency Trade-off}
\label{sec:tradeoff}

We consider three scoring strategies: independent scoring,
separate context-aware scoring, and joint context-aware
scoring. Let $f_\theta(q,C;u_i)$ denote the score for
candidate $u_i$ given question $q$ and context $C$,
and let $F_\theta(q,U)$ return scores for all candidates.
The three strategies are defined as
\begin{equation}
\begin{alignedat}{2}
r_i^{\mathrm{ind}}
    &= f_\theta(q,\{u_i\};u_i)
    &\quad& \text{\scriptsize Independent},\\
r_i^{\mathrm{sep}}
    &= f_\theta(q,U;u_i)
    && \text{\scriptsize Separate},\\
\mathbf{r}^{\mathrm{joint}}
    &= F_\theta(q,U)
    && \text{\scriptsize Joint}.
\end{alignedat}
\label{eq:scoring_strategies}
\end{equation}

Independent scoring considers only the target candidate.
Separate context-aware scoring considers the full candidate
set but requires $K$ inferences. Joint context-aware scoring
instead produces all $K$ scores through shared inference,
provided the candidate set fits within the model's
context window. Off-the-shelf rerankers can also be used
for independent scoring by ranking individual spans,
followed by the same budgeted selection procedure.
We compare these approaches in Table~\ref{tab:qampari_evidence}.

Using the same pretrained decision model, we observe
that separate context-aware scoring achieves higher
evidence recall than the other two strategies.
However, joint context-aware scoring requires
substantially less computation, at the cost of
lower decision quality. Detailed comparisons are
provided in Appendix~\ref{app:motivation_tradeoff}.

\subsection{Limitations of Pretrained Joint Decisions}
\label{sec:joint_scalability}

We examine whether pretrained joint decision models can
reliably identify supporting evidence as the number of
simultaneous decisions increases. We measure evidence
discrimination using AUROC, which evaluates how well
the model distinguishes supporting evidence from
distractor spans, independently of the context budget.

\begin{figure}[t]
    \centering
    \includegraphics[width=\columnwidth]
    {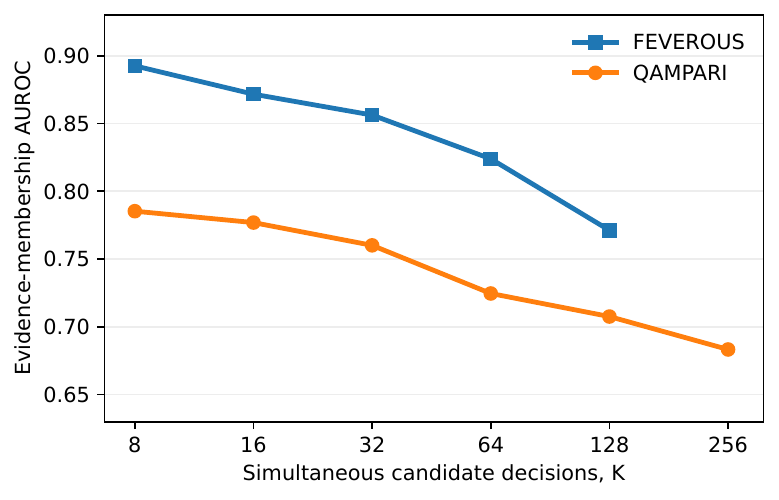}
    \caption{\textbf{Scaling limitations of pretrained joint decisions.}
    Evidence-discrimination AUROC decreases as the number
    of simultaneous decisions increases on multi-answer
    question answering and fact verification tasks.
    Higher is better.}
    \label{fig:joint_scalability}
\end{figure}

Figure~\ref{fig:joint_scalability} shows that the pretrained
model becomes less effective at distinguishing supporting
evidence as the number of candidate decisions increases.
We observe a consistent decline in AUROC on both
multi-answer question answering and fact verification
tasks. This suggests that the ability to produce many
decisions jointly does not guarantee their reliability.

One possible explanation is that larger candidate sets
also increase the input context length. However, our diagnostic experiments show that adding irrelevant text without increasing the number of decisions has a much smaller effect on discrimination. This suggests that
increasing the number of decisions is primarily responsible for the degradation. Detailed analysis is provided in
Appendix~\ref{app:decision_length}.

These observations reveal that the pre-trained decision model can produce many decisions
efficiently, but their reliability deteriorates as the candidate set grows. This motivates training \textsc{CURATOR} to make reliable joint evidence decisions across variable-sized candidate sets.

\section{CURATOR: Scalable Joint Evidence Selection}
\label{sec:curator}

We introduce \textsc{CURATOR}, a framework for adapting pre-trained joint decision models to reliable evidence selection over large candidate sets of spans. As shown in
Section~\ref{sec:joint_scalability}, pretrained joint decision models become less reliable at distinguishing supporting evidence from the distractors as the number of candidate decisions grows. Our method \textsc{CURATOR} addresses this limitation
through fine-tuning a decision model with evidence selection supervision over variable-sized candidate sets. We also interleave general decision tasks
in the fine-tuning process to reduce the risk of forgetting the pretrained
model's original decision-making capabilities.

\subsection{Constructing Joint Evidence Selection Data}
\label{sec:data_construction}

Fine-tuning for our evidence-selection task requires
a dataset with annotations identifying gold evidence
spans within passages that support the corresponding
answers.
We use three datasets that provide such annotations:
QAMPARI~\cite{amouyal-etal-2023-qampari},
HotpotQA~\cite{yang-etal-2018-hotpotqa}, and
2WikiMultiHopQA~\cite{ho-etal-2020-constructing}.
These datasets contain questions, supporting passages,
and the locations of supporting evidence within those
passages. In particular, QAMPARI contains questions
with multiple correct answers, each supported by
evidence distributed across multiple passages.

\paragraph{Evidence labels.}
We divide the passages into candidate spans and use the
dataset annotations to identify supporting evidence.
HotpotQA and 2WikiMultiHopQA provide supporting-fact
locations directly, while QAMPARI provides proof text
that we match to spans in the corresponding passages.
Supporting spans are labeled positive ($y_i=1$),
and other spans are labeled negative ($y_i=0$).
When the evidence label cannot be determined reliably,
we assign $y_i=\bot$. These uncertain spans remain
visible in the shared context but are excluded from
the training loss.

\paragraph{Variable-sized candidate sets.}
To address the degradation in joint decision quality
as the number of candidates increases, we construct
training examples with varying candidate counts and
evidence densities. We consider
$K\in\{8,16,32,64,128\}$ candidate spans.
For each candidate count, we use three sampling
strategies:

\begin{itemize}[leftmargin=*, nosep]
    \item \textbf{Natural sampling.}
    We uniformly sample the desired number of candidate
    spans. This preserves the naturally sparse
    distribution of evidence within the text.
    This sampling can also produce candidate sets
    with no supporting evidence.

    \item \textbf{Fixed-positive sampling.}
    We select exactly three supporting spans and
    fill the remaining $K-3$ positions with distractors.
    As $K$ increases, the supporting evidence
    becomes increasingly sparse.

    \item \textbf{Gold-included sampling.}
    We include up to $K-1$ annotated supporting spans
    and fill the remaining positions with distractors.
    This produces evidence-rich examples requiring
    multiple positive decisions within the same context.
\end{itemize}

These strategies address two challenges of joint
evidence selection. Varying $K$ trains the model to
distinguish relevant spans as the number of simultaneous
decisions increases. Varying evidence density exposes
the model to both sparse and evidence-rich contexts,
rather than restricting training to predominantly
negative candidate sets. Together, they aim to improve
evidence discrimination across different joint
selection workloads.

\subsection{Joint Decision Model}
\label{sec:joint_model}

We develop our model \textsc{CURATOR} on
\texttt{pplx-decider-flash-v1.2}, a pretrained
joint decision model based on Qwen3.5-4B
\cite{perplexity2026deciderflash}. 
The model supports multiple decision readouts
within a shared input context. This allows us to score several candidates without generating
a separate token response for each decision.

Given a question $q$ and candidate set
$U=(u_1,\ldots,u_K)$, we associate each candidate $u_i$ with a binary decision asking whether it is an supporting evidence to answer $q$. The decision model accordingly produces positive and negative logits,
$\ell_i^{(1)}$ and $\ell_i^{(0)}$, for each candidate.
We define its evidence score and positive-class
probability as
\begin{equation}
\begin{aligned}
    r_i &= \ell_i^{(1)}-\ell_i^{(0)},\\
    p_i &= \sigma(r_i/T),
\end{aligned}
\label{eq:evidence_margin}
\end{equation}
where $T$ is the decision model's temperature
and $\sigma$ is the sigmoid function.
A higher value of $r_i$ suggests stronger evidence relevance for the candidate $u_i$.

When the entire candidate set fits
within the model's input window, all the $K$
scores are produced in a single inference. However, when the candidate pool is larger, it is divided into smaller groups to fit the context window. The candidates are scored within each group, and the resulting scores are merged. Finally, we rank the candidates based on the score and select them within the given context budget for retrieval (see Equation~\ref{eq:evidence_budget}).

\paragraph{Parameter-efficient adaptation.}
The pretrained model combines causal
linear-attention layers with noncausal
full-attention layers in transformers. We adapt the query,
key, value, and output projections of its
eight full-attention layers using low-rank
adaptation (LoRA), with rank 16 and scaling
parameter 32. We also train the existing
decision readout while keeping the remaining
backbone parameters frozen. Following this we only update approximately 3.8 million
parameters out of the 4.54-billion-parameter
model, allowing us to specialize its joint
evidence decisions without full-model
fine-tuning.

\subsection{Class-Balanced Joint Evidence Supervision}
\label{sec:evidence_objective}

A central challenge in evidence selection is
that the actual supporting spans may constitute a small fraction of the whole candidate set. For example, with three supporting sentences,
increasing $K$ from 8 to 128 reduces the
positive proportion from 37.5\% to 2.3\%.
Under ordinary binary cross-entropy loss, this
large number of negative decisions can
dominate the training objective.

We address this imbalance using masked,
class-balanced supervision over the joint
decision readouts. For each labeled class
$c\in\{0,1\}$, let
$\mathcal{I}_c=\{i:y_i=c\}$ denote its
candidate indices, and let
$\mathcal{C}=\{c:|\mathcal{I}_c|>0\}$
denote the classes present in the request.
Using $p_i^{(1)}=p_i$ and
$p_i^{(0)}=1-p_i$, we define
\begin{equation}
\label{eq:joint_evidence_loss}
\mathcal{L}_{\mathrm{evidence}}
=
-\sum_{c\in\mathcal{C}}
\frac{1}{|\mathcal{I}_c|}
\sum_{i\in\mathcal{I}_c}
\log p_i^{(c)}.
\end{equation}

In our training, we skip supervision for candidates with uncertain labels ($y_i=\bot$). 

Although the loss supervises individual
readouts, all decisions are made from the
shared candidate context and update the
same model parameters. The model therefore
learns to identify supporting evidence in
the presence of other candidates, rather
than learning only to classify isolated
sentences. Combined with variable-sized
training sets, this supervision directly
targets the deterioration in joint evidence
discrimination observed in
Section~\ref{sec:joint_scalability}.

\subsection{Preserving General Decision Capabilities}
\label{sec:general_replay}

Adapting a pretrained decision model for evidence selection
may degrade its original decision-making capabilities.
To mitigate this, we use general decision-task replay
\cite{buzzega2020dark} with examples covering binary
classification, multiple-choice selection, and graded
scoring. These tasks cover different decision /classification formats
supported by the pretrained decision model
\cite{sileo-2024-tasksource,perplexity2026deciderflash}.
Since the model's complete original training data is not
publicly available, we use labeled public benchmarks
for replay.

We combine supervision from ground-truth labels with
knowledge distillation from the frozen pretrained model:
\begin{equation}
\begin{aligned}
\mathcal{L}_{\mathrm{replay}}
={}&
\lambda_{\mathrm{CE}}
\left[-\log P_\theta^\tau(y\mid x)\right]\\
&+
\lambda_{\mathrm{KD}}\tau^2
D_{\mathrm{KL}}
\left(
P_0^\tau(\cdot\mid x)
\,\|\,
P_\theta^\tau(\cdot\mid x)
\right),
\end{aligned}
\label{eq:replay_objective}
\end{equation}
where $(x,y)$ is a labeled replay example, and
$P_0^\tau$ and $P_\theta^\tau$ denote the frozen
pretrained and adapted model distributions at
temperature $\tau$. We set
$\lambda_{\mathrm{CE}}=\lambda_{\mathrm{KD}}=0.5$.

The overall training loss averages the evidence and
replay objectives over their respective examples:
\begin{equation}
\mathcal{L}_{\mathrm{train}}
=
\frac{1}{|\mathcal{B}|}
\sum_{x\in\mathcal{B}}
\mathcal{L}(x),
\label{eq:training_objective}
\end{equation}
where
\begin{equation}
\mathcal{L}(x)=
\begin{cases}
\mathcal{L}_{\mathrm{evidence}}(x),
    & x\in\mathcal{B}_{\mathrm{evidence}},\\
\mathcal{L}_{\mathrm{replay}}(x),
    & x\in\mathcal{B}_{\mathrm{replay}}.
\end{cases}
\end{equation}
We sample 60\% of training examples from joint
evidence-selection tasks and 40\% from general
decision tasks.

Additional data-construction details, model implementation,
training objectives, and ablation analyses are provided in
Appendix~\ref{app:curator_details}.

\section{Experiments}
\label{sec:experiments}

We conduct experiments covering four broad research questions (RQ) that highlight the generalization capability, efficiency, and downstream RAG performance of \textsc{CURATOR}.

\begin{itemize}[leftmargin=*, nosep]
    \item \textbf{RQ1: Evidence selection and efficiency (Section~\ref{sec:evidence_quality}).}
    In a fixed token budget setting, is it possible to recover as much relevant evidence as possible while retaining the computational efficiency of joint inference? To answer this, we measure evidence recall, scoring latency, and input-token cost on QAMPARI.

    \item \textbf{RQ2: Generalization Capability (Section~\ref{sec:generalization}).}
    Is our method generalizable to unseen domains and evidence-selection tasks? We evaluate evidence recall and complete-set coverage on datasets used for fact verification (FEVEROUS), financial documents (MultiHiertt), and legal contracts (CUAD).

    \item \textbf{RQ3: Downstream performance in a RAG setting(Section~\ref{sec:downstream_rag}).}
    In a budget-constrained Retrieval-Augmented Generation (RAG) setting, does better evidence retrieval improve the end-to-end answer quality? Using a 120B parameter LLM for generation in a RAG setup, we evaluate normalized exact match scores under different context budgets on MuSiQue and MultiHiertt.

    \item \textbf{RQ4: Joint-decision scalability (Section~\ref{sec:scalability_results}).}
    Can CURATOR maintain performance as the number of simultaneous decisions increase? We measure and report AUROC on FEVEROUS and QAMPARI using varying sizes of controlled candidate sets.

\end{itemize}

\subsection{Experimental Setup}
\label{sec:experimental_setup}

\paragraph{Evaluation datasets.}
We evaluate evidence selection across multi-answer QA \cite{amouyal-etal-2023-qampari},
fact verification \cite{aly-etal-2021-fact}, financial documents \cite{hendrycks2021cuadexpertannotatednlpdataset},
and legal contracts \cite{zhao-etal-2022-multihiertt}. We additionally evaluate on downstream answer generation on MuSiQue \cite{trivedi-etal-2022-musique} and MultiHiertt \cite{zhao-etal-2022-multihiertt}.
Table~\ref{tab:dataset_statistics} summarizes the
evaluation datasets and their evidence characteristics.

\begin{table}[t]
\centering
\scriptsize
\setlength{\tabcolsep}{2pt}
\caption{\textbf{Evaluation dataset statistics.}
Candidate and gold counts are means per question.
Density is the mean per-question fraction of candidate
units containing annotated evidence. Pool coverage is
the mean fraction of annotated evidence available before
selection. FEVEROUS includes annotated evidence pages
and eight BM25 distractor pages per claim; its full
coverage is by construction.}
\label{tab:dataset_statistics}
\resizebox{\columnwidth}{!}{%
\begin{tabular}{@{}lrrrrr@{}}
\hline
\textbf{Dataset}
& \textbf{Questions}
& \textbf{Cand./q}
& \textbf{Gold/q}
& \textbf{Density}
& \textbf{Pool cov.} \\
\hline
QAMPARI     & 855   & 56.51 & 15.31 & 18.13\% & 76.13\% \\
FEVEROUS    & 6,808 & 343.0 & 3.94  & 1.29\%  & 100.00\% \\
MuSiQue     & 2,417 & 75.5  & 10.46 & 13.91\% & 100.00\% \\
MultiHiertt & 1,044 & 119.0 & 3.37  & 3.62\%  & 93.89\% \\
CUAD        & 525   & 166.2 & 2.26  & 1.41\%  & 96.55\% \\
\hline
\end{tabular}%
}
\end{table}

\paragraph{Models and baselines.}
We compare the following evidence-selection methods:

\begin{itemize}[leftmargin=*, nosep]
    \item \textbf{Original Joint Decider.}
    The pretrained \texttt{pplx-decider-flash-v1.2}
    model~\cite{perplexity2026deciderflash},
    before evidence-specific adaptation.

    \item \textbf{Span-level Rerankers.}
    Qwen3-Reranker-4B~\cite{zhang2025qwen3embeddingadvancingtext}
    and BGE-reranker-v2-gemma~
    \cite{bge_reranker_v2_gemma,chen-etal-2024-m3},
    applied to individual evidence spans. Generally, rerankers are used to rank chunks/passage but to get a fair comparison with our method, we also use them to score spans.

    \item \textbf{Passage/Chunk-level Reranker.}
    Qwen3-Reranker-4B applied to larger retrieved
    chunks rather than individual spans. Exactly how it is supposed to be applied.

    \item \textbf{CURATOR (Ours).}
    The adapted joint decision model described
    in Section~\ref{sec:curator}.
\end{itemize}

We also evaluate separate context-aware scoring,
defined in Section~\ref{sec:tradeoff} (Eqn ~\ref{eq:scoring_strategies}), using the
original pretrained decision model.

\paragraph{Evaluation candidate pools.}
Within each experiment, methods use the same retrieved
source pool. For QAMPARI, we divide the top-30 retrieved
passages into spans of up to three consecutive sentences.
For FEVEROUS, we combine annotated evidence pages with
eight BM25 distractor pages and extract sentence-level
and table-cell units as spans. MuSiQue and CUAD use sentence-level
units, while MultiHiertt uses sentences and table cells.
Downstream RAG evaluation uses up to 20 retrieved passages
per question. Details of candidate construction and
large-pool scoring are provided in
Appendix~\ref{app:curator_inference}.

\paragraph{Context budgets and metrics.}
We evaluate applicable context budgets
$B\in\{64,128,256,512,1024,2048\}$ tokens.
\textbf{Evidence Recall} measures the fraction of annotated supporting units selected, divided by the number of gold evidences in the denominator. \textbf{Complete-Set Coverage} indicates the
fraction of questions for which all of the possible gold evidence has been selected. FEVEROUS reports pooled gold-unit recall, while other datasets use mean per-question recall. In conjunction with this, we report \textbf{AUROC} for evidence discrimination at scale,
\textbf{Normalized Answer Match} for downstream RAG generation. For demonstrating efficiency, we report
\textbf{scoring latency} and \textbf{input-token consumption}.

\subsection{Evidence Selection Quality and Efficiency}
\label{sec:evidence_quality}

We evaluate whether CURATOR improves context-aware evidence selection without repeatedly processing the retrieved context for each candidate. Table~\ref{tab:qampari_evidence} reports evidence recall across three context budgets on 855 QAMPARI development questions, alongside the computational cost of scoring the candidate pool.

\begin{table*}[t]
\centering
\small
\setlength{\tabcolsep}{6pt}
\caption{\textbf{Evidence selection quality and efficiency on QAMPARI.}
Evidence recall is reported at three selected-context budgets.
Scoring latency is averaged per question, while input tokens are median totals per question. Candidate scoring is performed once, independently of the selected-context budget. Results are from the development set, which overlaps with checkpoint-selection data.}
\label{tab:qampari_evidence}
\begin{tabular}{lccccc}
\hline
& \multicolumn{3}{c}{\textbf{Evidence Recall} $\uparrow$}
& \multicolumn{2}{c}{\textbf{Scoring Cost} $\downarrow$} \\
\textbf{Method}
& $B=512$ & $B=1024$ & $B=2048$
& Time (s/q) & Input tokens/q \\
\hline
Original Joint Decider
& 0.2763 & 0.4279 & 0.5791 & 0.909 & 11,411 \\
Separate Context-Aware
& 0.3166 & 0.4844 & 0.6242 & 20.065 & 293,296 \\
Qwen3-Reranker (Span)
& 0.1874 & 0.3316 & 0.5230 & 0.688 & 6,664 \\
BGE-Reranker (Span)
& 0.2297 & 0.3757 & 0.5462 & 0.301 & 7,046 \\
Qwen3-Reranker (Chunk)
& 0.1827 & 0.3042 & 0.4815 & 0.449 & 5,396 \\
\textbf{CURATOR (Ours)}
& \textbf{0.3663} & \textbf{0.5466} & \textbf{0.6784}
& 0.925 & 11,411 \\
\hline
\end{tabular}
\end{table*}

\paragraph{Accuracy--efficiency trade-off.}
Separate context-aware scoring achieves higher recall than the original joint decider, suggesting that the pretrained model makes more accurate decisions when evaluating one candidate at a time. However, this requires repeated full-context inference. CURATOR improves joint decision quality, achieving 17--33\% higher recall than the original joint decider and 9--16\% higher recall than separate context-aware scoring across the three budgets. Moreover, CURATOR is $21.7\times$ faster and uses $25.7\times$ fewer scoring-input tokens than separate context-aware scoring, while maintaining the computational cost of the original joint inference.

\paragraph{Comparison with specialized rerankers.}
CURATOR also outperforms specialized relevance-ranking models. At $B=1024$, it achieves 64.8\% and 45.5\% higher recall than Qwen3-Reranker and BGE-Reranker at span level, respectively. Span-level Qwen3-Reranker consistently outperforms its chunk-level counterpart, highlighting the advantage of fine-grained evidence selection. Although these rerankers have lower scoring latency, CURATOR achieves substantially higher evidence recall.

\paragraph{Impact of context budget.}
CURATOR's advantage is largest under tighter context budgets. Its relative gain over separate context-aware scoring increases from 8.7\% at $B=2048$ to 15.7\% at $B=512$, while its gain over the original joint decider increases from 17.1\% to 32.6\%. This suggests that accurate evidence selection is particularly valuable when the available context is limited. Complementing this result, Section~\ref{sec:scalability_results} shows that CURATOR also maintains stronger evidence discrimination as the number of simultaneous decisions increases.

\subsection{Generalization Across Tasks and Domains}
\label{sec:generalization}
\begin{table*}[t]
\centering
\small
\setlength{\tabcolsep}{4pt}
\caption{\textbf{Generalization of evidence selection across tasks and domains.}
Evidence recall measures the fraction of annotated supporting evidence retained, while complete-set coverage requires all annotated gold units for a question to be selected.}
\label{tab:cross_domain}
\begin{tabular}{llrrrrrr}
\hline
& & \multicolumn{3}{c}{\textbf{Evidence Recall (\%)} $\uparrow$}
& \multicolumn{3}{c}{\textbf{Complete-Set Coverage (\%)} $\uparrow$} \\
\textbf{Dataset} & \textbf{Method}
& $B=128$ & $B=512$ & $B=2048$
& $B=128$ & $B=512$ & $B=2048$ \\
\hline
FEVEROUS
& Original Joint Decider
& 33.4 & 57.0 & 70.8
& 12.9 & 33.5 & 47.7 \\
& Qwen3-Reranker (Span)
& 40.0 & 74.6 & 91.3
& 18.3 & 55.3 & 79.9 \\
& \textbf{CURATOR}
& \textbf{50.4} & \textbf{84.7} & \textbf{96.2}
& \textbf{26.8} & \textbf{70.4} & \textbf{90.4} \\
\hline
MultiHiertt
& Original Joint Decider
& 53.3 & 68.2 & 82.2
& 23.0 & 34.3 & 54.5 \\
& Qwen3-Reranker (Span)
& 57.9 & 80.6 & 91.2
& 27.1 & 50.6 & 72.9 \\
& \textbf{CURATOR}
& \textbf{64.1} & \textbf{84.5} & \textbf{92.5}
& \textbf{31.7} & \textbf{58.3} & \textbf{76.3} \\
\hline
CUAD
& Original Joint Decider
& 59.4 & 78.0 & 89.6
& 49.3 & 66.7 & 81.0 \\
& Qwen3-Reranker (Span)
& 52.2 & 78.5 & 91.8
& 41.7 & 67.6 & 84.6 \\
& \textbf{CURATOR}
& \textbf{68.0} & \textbf{86.7} & \textbf{94.4}
& \textbf{57.5} & \textbf{79.8} & \textbf{89.9} \\
\hline
\end{tabular}
\end{table*}
We next examine whether CURATOR can identify supporting evidence in tasks and document domains not used for evidence-selection training. We evaluate FEVEROUS on fact verification, MultiHiertt on financial reports, and CUAD on legal contracts. The latter two differ substantially from the Wikipedia-based training data. Table~\ref{tab:cross_domain} compares evidence recall and complete-set coverage at three representative context budgets.

\textbf{Generalization to fact verification.}
On FEVEROUS, CURATOR improves evidence selection despite the sparse supporting evidence and additional BM25 distractors. At $B=512$, CURATOR achieves 48.6\% higher recall than the original joint decider, while complete-set coverage more than doubles. These gains persist across all evaluated budgets, demonstrating transfer to fact verification involving both textual and tabular evidence.

\textbf{Generalization across domains.}
CURATOR also improves evidence recall on financial reports and legal contracts, despite neither domain contributing to its evidence-selection training. At $B=2048$, CURATOR achieves 1.5\% higher recall than the strongest baseline on MultiHiertt and 2.8\% on CUAD. As the context budget decreases to $B=128$, these improvements increase to 10.7\% and 14.4\%, respectively. This suggests that the benefits of evidence selection transfer across domains and are particularly valuable under restrictive context budgets.


\subsection{Impact on Downstream RAG Performance}
\label{sec:downstream_rag}

We next examine whether improved evidence selection translates into better downstream answers. We evaluate MuSiQue, MultiHiertt using 120B sized frontier LLM. Within each dataset, all methods use the same retrieved passages and answering model, differing only in the selected context.

\begin{table}[t]
\centering
\footnotesize
\setlength{\tabcolsep}{3pt}
\caption{\textbf{Downstream answer quality.}
Normalized answer match (\%) across context
budgets (tokens). Higher is better.}
\label{tab:downstream_rag}

\begin{tabular*}{\columnwidth}
{@{\extracolsep{\fill}}lrrrr@{}}
\hline
\textbf{Method}
& \multicolumn{4}{c}{\textbf{Context budget (tokens)}} \\
& \textbf{128} & \textbf{512}
& \textbf{1024} & \textbf{2048} \\
\hline
\multicolumn{5}{c}{\textbf{MuSiQue}} \\
\hline
Qwen3 (Chunk)
& 19.3 & 47.0 & 53.0 & 56.6 \\
Qwen3 (Span)
& 39.7 & 52.4 & 53.7 & 56.5 \\
Original Joint
& 40.3 & 47.1 & 48.7 & 56.6 \\
\textbf{CURATOR}
& \textbf{48.7} & \textbf{55.7}
& \textbf{55.0} & \textbf{56.8} \\
\hline
\multicolumn{5}{c}{\textbf{MultiHiertt}} \\
\hline
Qwen3 (Chunk)
& 9.0 & 13.8 & 22.7 & 37.5 \\
Qwen3 (Span)
& 42.4 & 53.2 & 54.7 & \textbf{55.2} \\
Original Joint
& 38.3 & 48.0 & 49.5 & 53.5 \\
\textbf{CURATOR}
& \textbf{48.4} & \textbf{55.5}
& \textbf{55.8} & 55.1 \\
\hline
\end{tabular*}
\end{table}

\paragraph{Improving downstream answer quality.}
CURATOR achieves the best performance on both datasets at budgets between 128 and 1,024 tokens. At $B=256$, it improves answer accuracy over the strongest baseline by 20.4\% on MuSiQue and 9.1\% on MultiHiertt. These gains demonstrate that improved evidence selection translates into better downstream generation across different tasks and answering models.

\paragraph{Effect of context budget.}
CURATOR's advantage is strongest under constrained budgets. At $B=128$, its relative improvements over the strongest baselines reach 20.9\% on MuSiQue and 14.0\% on MultiHiertt, decreasing to 6.4\% and 4.3\% at $B=512$. CURATOR also approaches or slightly exceeds the unfiltered whole-pool reference using substantially less context, demonstrating the benefit of selective evidence admission.

\paragraph{Evidence quality versus answer accuracy.}
Better evidence recovery does not translate proportionally into downstream performance. At $B=512$ on MuSiQue, CURATOR improves complete-set coverage of annotated supporting sentences by approximately $31\times$ over the original joint decider, while answer accuracy improves by 18.3\%. This difference highlights the role of the answering model in effectively using the selected evidence.

\subsection{Scalability of Joint Evidence Decisions}
\label{sec:scalability_results}

We examine whether CURATOR addresses the decline in joint decision quality identified in Section~\ref{sec:joint_scalability}. To evaluate evidence discrimination independently of retrieval coverage, we construct controlled pools from QAMPARI and FEVEROUS, each containing three supporting units and $K-3$ distractors. Pools are sampled independently at each $K$, and both models score identical candidates in a single joint request without truncation. We measure pooled evidence-membership AUROC as $K$ increases.

\begin{figure*}[t]
    \centering
    \includegraphics[width=\textwidth]{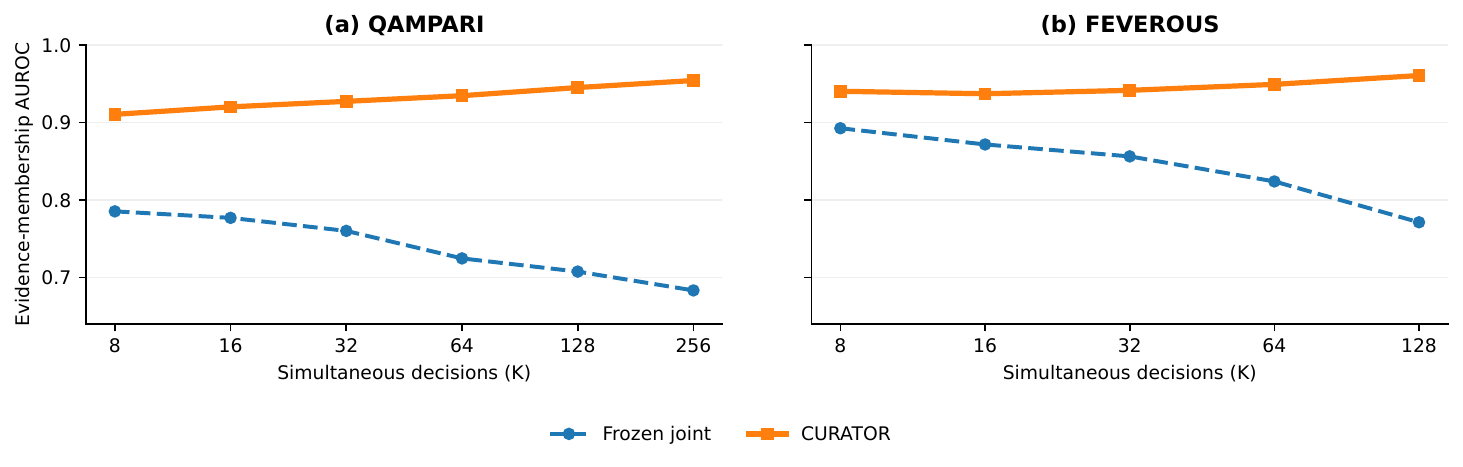}
    \caption{\textbf{Scalability of joint evidence decisions.} Pooled evidence-membership AUROC for the Original Joint Decider and CURATOR on (a) QAMPARI and (b) FEVEROUS as the number of simultaneous decisions $K$ increases. Each controlled pool contains three supporting units and $K-3$ distractors, independently sampled at each $K$. Both models evaluate the same pools without truncation. Higher is better.}
    \label{fig:auroc_scaling_results}
\end{figure*}

\paragraph{Scaling to larger candidate sets.}
Figure~\ref{fig:auroc_scaling_results} shows that the original joint decider loses discrimination as $K$ grows, while CURATOR maintains strong performance. The AUROC gap widens monotonically from $0.125$ to $0.271$ on QAMPARI ($K=8$ to $256$), and from $0.048$ to $0.189$ on FEVEROUS ($K=8$ to $128$). CURATOR's advantage thus grows with the joint decision workload.

The improvement transfers to FEVEROUS, which was not used for evidence-selection training, and extends to $K=256$ on QAMPARI, beyond the largest training candidate count ($K\leq128$). Because pools are independently sampled across $K$, these results demonstrate stronger large-set evidence discrimination without isolating decision count as the sole cause.

\subsection{Ablations and Preservation of General Capabilities}
\label{sec:ablations_retention}

\paragraph{Ablation studies.}
We investigate which training choices contribute to CURATOR's improved joint decision scalability. Pointwise evidence adaptation improves discrimination but retains the original model's degradation as the candidate count grows. In contrast, joint training with small candidate sets largely eliminates this decline, while variable-$K$ training further improves large-set discrimination. These results highlight the importance of joint evidence supervision. Detailed results are provided in Appendix~\ref{app:ablations}.

\paragraph{Preservation of general capabilities.}
We evaluate whether evidence-specific adaptation affects the pretrained model's original decision capabilities. Across seven general decision benchmarks covering 19,631 examples, CURATOR achieves a macro-average of 0.8810 compared with 0.8733 for the original model. No individual task shows a statistically detectable difference, indicating no detectable degradation in the measured capabilities. Detailed per-task results are provided in Appendix~\ref{app:general_retention}.

\section{Related Work}
\label{sec:related_work}

\paragraph{Fine-grained retrieval and context compression.}
Retrieval-augmented generation (RAG) improves language models by incorporating retrieved information \cite{3495724.3496517,3524938.3525306,izacard-grave-2021-leveraging}. Conventional rerankers assess the relevance of retrieved passages or spans, typically scoring each candidate separately against the query \cite{nogueira2020passagererankingbert,nogueira-etal-2020-document,Khattab2020ColBERTEA,zhang2025qwen3embeddingadvancingtext}. Listwise reranking instead considers multiple candidates together, but often relies on repeated generation to produce rankings \cite{sun-etal-2023-chatgpt,pradeep2023rankzephyreffectiverobustzeroshot}. Related work on context compression reduces unnecessary retrieved information through extractive selection or token-level pruning \cite{ICLR2024_bda88ed2,jiang-etal-2023-llmlingua,jiang-etal-2024-longllmlingua,li-etal-2023-compressing,hwang-etal-2025-exit}. Unlike these approaches, we focus on making context-aware decisions for multiple evidence spans jointly, selecting original text under a fixed token budget.

\paragraph{Decision models for joint scoring.}
Decision models score predefined choices directly, avoiding the need to generate textual judgments \cite{li2026llmasjevllmsjevstyledecision,geuter2026releasinginnerjevsimple}. Recent multi-readout models extend this capability to multiple decisions within a shared inference \cite{perplexity2026deciderflash,perplexity2026decider}. This makes them suitable for evidence selection, where the relevance of one span may depend on information in other retrieved passages. However, we find that pretrained decision models lose discrimination as the number of simultaneous decisions increases. We address this limitation by adapting a pretrained decision model through joint evidence supervision over variable-sized candidate sets.

\paragraph{Related decision-based retrieval.}
Concurrent work on DecisionRAG \cite{wan2026retrieverfreeretrievalnavigatingdocuments} also uses parallel decisions from a pretrained decision model and observes limitations when evaluating large candidate sets. DecisionRAG retains a frozen model and uses hierarchical retrieval to divide large candidate pools into smaller groups. In contrast, \textsc{CURATOR} improves the model's ability to make many decisions jointly through training, while operating on a fixed retrieved pool. These approaches are complementary: improving joint decision capacity could also benefit hierarchical retrieval. Unlike DecisionRAG, we evaluate fine-grained evidence selection under token budgets and its impact on downstream RAG performance. The two methods use different evaluation settings, so we do not make a direct performance comparison.

\section{Conclusion}
\label{sec:conclusion}

We introduced \textsc{CURATOR}, a span-level evidence selection framework that adapts a pretrained joint decision model to identify relevant information from retrieved documents under a limited context budget. By training the model to make multiple evidence decisions jointly, CURATOR improves selection quality while retaining the computational efficiency of shared inference. Our experiments demonstrate consistent improvements in evidence recall and complete-set coverage, particularly under tight token budgets. These improvements generalize to fact verification, financial documents, and legal contracts, and translate into better downstream RAG performance. Controlled scaling experiments further show that CURATOR maintains strong evidence discrimination as the number of candidates increases, addressing a key limitation of pretrained joint decision models.

Our results highlight the potential of decision models for fine-grained context selection, moving beyond passage-level retrieval toward more selective and efficient use of retrieved information. CURATOR's joint scoring remains reliable at large candidate counts, suggesting it could extend naturally to agentic settings by re-ranking the accumulated evidence pool as new tool outputs arrive, dynamically sharing the token budget between retrieved evidence and the agent's working context. We leave this multi-step setting to future work. 
\section*{Limitations}

\section*{Acknowledgments}
We used generative AI tools to assist with improving the clarity and presentation of the manuscript, generating illustrative figures, and formatting. All AI-assisted content was reviewed and verified by the authors, who take full responsibility for the final manuscript and its scientific claims.


\bibliography{custom}

\appendix

\section{Diagnostics of the Decision Models}
\label{app:motivation_diagnostics}

\subsection{Accuracy--Efficiency Trade-off of Frozen Scoring Strategies}
\label{app:motivation_tradeoff}

We compare independent scoring, separate context-aware scoring,
and joint context-aware scoring using the same frozen pretrained
decision model. All three strategies use identical candidate pools,
scoring definitions, and budgeted selection procedures within
each dataset.

We evaluate matched 200-example panels from FEVEROUS and QAMPARI.
The FEVEROUS panel uses annotated evidence pages with a 120-unit
candidate cap and is distinct from the full distractor-augmented
evaluation in Section~\ref{sec:generalization}. The QAMPARI panel
is likewise a smaller diagnostic sample rather than the full
development evaluation.

\begin{table}[t]
    \centering
    \small
    \setlength{\tabcolsep}{4pt}
    \caption{\textbf{Accuracy--efficiency trade-off of frozen
    evidence decisions.} Evidence recall at a selected-context
    budget of 512 tokens for FEVEROUS and 1,024 tokens for
    QAMPARI, alongside mean scoring latency per question.
    Each dataset contains 200 examples.}
    \label{tab:motivation_tradeoff}
    \begin{tabular}{llrr}
        \hline
        Dataset & Scoring strategy & Recall $\uparrow$ & Time (s) $\downarrow$ \\
        \hline
        FEVEROUS
        & Independent & .743 & 1.48 \\
        & Separate context-aware & \textbf{.814} & 14.93 \\
        & Joint context-aware & .676 & \textbf{.85} \\
        \hline
        QAMPARI
        & Independent & .428 & 1.57 \\
        & Separate context-aware & \textbf{.504} & 20.25 \\
        & Joint context-aware & .460 & \textbf{.91} \\
        \hline
    \end{tabular}
\end{table}

Separate context-aware scoring achieves the highest recall on
both panels, showing that access to other candidate evidence
can improve individual judgments. However, the pretrained
model does not consistently preserve this advantage when
making all decisions jointly. Joint context-aware scoring
performs below independent scoring on FEVEROUS and remains
below separate context-aware scoring on QAMPARI.

The computational difference is substantial. On QAMPARI,
joint scoring reduces latency by approximately $22\times$
compared with separate context-aware scoring and processes
11,408 median input tokens per question rather than 293,227.
On FEVEROUS, the corresponding latency reduction is
approximately $18\times$. These measurements use matched
single-GPU inference without cross-question batching or
caching. Thus, joint inference reduces repeated computation,
but the quality of pretrained joint evidence decisions
remains a limitation.

\subsection{Evidence Discrimination with Increasing Candidate Sets}
\label{app:frozen_scalability}

To examine how joint decision quality changes with candidate
set size, we evaluate the frozen \texttt{pplx-decider-flash-v1.2}
model on controlled candidate pools from QAMPARI and FEVEROUS.
Each pool contains three annotated supporting units and
$K-3$ distractors. The supporting units are included by
construction, allowing us to measure evidence discrimination
independently of whether the retrieval system found them.

For QAMPARI, we evaluate 424 questions with two independently
sampled candidate sets per question at each $K$, covering
$K\in\{8,16,32,64,128,256\}$. For FEVEROUS, we evaluate
1,000 questions with two sampled sets per question, covering
$K\in\{8,16,32,64,128\}$. Candidate pools are sampled
independently at each $K$ and are not nested. Each pool
is scored in a single joint inference without truncation.

We compute evidence-membership scores using the difference
between the positive and negative decision logits. We
report pooled AUROC across the candidate scores and labels
at each $K$. AUROC measures the ability to rank supporting
evidence above distractors without depending on a particular
selection threshold or token budget.

As shown in Figure~\ref{fig:joint_scalability}, the pretrained
model's discrimination deteriorates as the candidate set
grows. On QAMPARI, AUROC decreases from 0.785 at $K=8$
to 0.683 at $K=256$. On FEVEROUS, it decreases from
0.893 at $K=8$ to 0.771 at $K=128$.

These results reveal a scaling limitation of pretrained
joint evidence decisions. However, increasing $K$ also
changes the distractor composition and input length.
The observed decline therefore cannot be attributed
exclusively to the number of requested decisions.

\subsection{Effect of Decision Count and Input Length}
\label{app:decision_length}

The deterioration in joint evidence discrimination may
result from increasing decision count, longer inputs,
or additional distractors. To examine the role of input
length, we conduct a diagnostic using 15 paired QAMPARI
questions, each with three supporting candidates.

We vary the candidate count $K$ from 8 to 128. At each
fixed $K$, we additionally append 6,000 or 14,000
irrelevant tokens to the shared context without introducing
new decisions. Every candidate set is evaluated in a
single joint request without truncation. We report
AUROC computed within each candidate set and averaged
across paired examples.

\begin{figure}[t]
    \centering
    \includegraphics[width=\columnwidth]
    {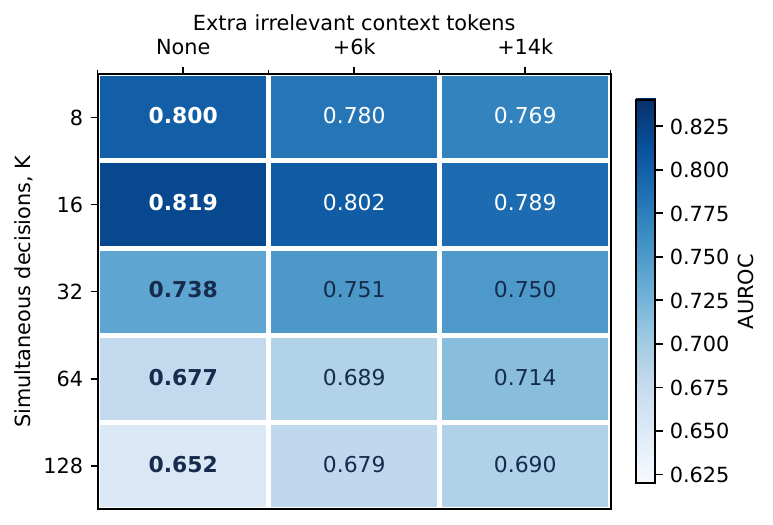}
    \caption{\textbf{Decision count versus added context length.}
    Mean within-set AUROC on 15 paired QAMPARI questions
    containing three supporting candidates. Rows vary
    the number of candidate decisions, while columns
    add irrelevant text without increasing the number
    of decisions. Higher is better.}
    \label{fig:decision_count_length}
\end{figure}

Without padding, AUROC decreases from 0.800 at $K=8$
to 0.652 at $K=128$. In contrast, adding irrelevant
text at a fixed $K$ produces smaller, non-monotonic
changes. At $K=128$, adding 14,000 irrelevant tokens
changes AUROC from 0.652 to 0.690 rather than causing
a further decrease.

These observations suggest that increasing input length
alone is insufficient to explain the deterioration.
However, the sample is small, and increasing $K$ also
changes the distractor composition. The experiment
therefore does not establish a causal effect of
decision count alone.

\section{CURATOR: Implementation Details and Additional Analyses}
\label{app:curator_details}

\subsection{Evidence Annotation and Candidate Construction}
\label{app:curator_data}

We construct evidence-selection training examples from
QAMPARI, HotpotQA, and 2WikiMultiHopQA. Each example
contains a question, a set of candidate sentences, and
evidence-membership annotations. The candidate sentences
are extracted from the passages associated with each
question.

The datasets provide different forms of evidence
annotations. HotpotQA and 2WikiMultiHopQA identify
supporting facts at the sentence level. For QAMPARI,
we align annotated proof text with sentences from
the corresponding passages. These annotations provide
the positive candidates used for training.

We distinguish three types of candidate labels:
positive, verified negative, and uncertain. Positive
candidates are identified by supporting-evidence
annotations. Negative labels are assigned only to
candidates treated as verified non-evidence by the
data-construction procedure. Candidates whose evidence
membership cannot be established are marked uncertain.

Uncertain candidates are retained in the model input
because they are part of the retrieved context. However,
their readouts are excluded from the training objective.
This distinction is important when evidence annotations
are incomplete: the absence of an annotation does not
necessarily mean that a sentence is irrelevant.

\paragraph{Sampling candidate sets.}
For each training question, we construct joint requests
with $K\in\{8,16,32,64,128\}$ candidate sentences.
The three sampling strategies described in
Section~\ref{sec:data_construction} expose the model
to different evidence distributions.

Fixed-positive sampling constructs requests containing
three supporting sentences and $K-3$ distractors.
For example, a request with $K=8$ contains 37.5\%
supporting evidence, whereas a request with $K=128$
contains only 2.3\%. This creates increasingly sparse
selection problems without changing the number of
supporting candidates.

Natural sampling draws from the underlying candidate
pool without requiring a fixed number of positives.
It therefore includes requests with varying evidence
density, including requests without annotated supporting
evidence.

Gold-included sampling prioritizes the annotated
supporting sentences, subject to the capacity $K$,
and fills the remaining positions with distractors.
Unlike fixed-positive sampling, this strategy allows
requests containing different numbers of supporting
sentences.

The combination of these strategies separates two
important sources of variation: the number of decisions
made jointly and the proportion of decisions that
correspond to supporting evidence.


\subsection{Model Adaptation and Training Configuration}
\label{app:curator_implementation}

We initialize \textsc{CURATOR} from
\texttt{pplx-decider-flash-v1.2}, a 4.54B-parameter
pretrained decision model. We retain its existing
decision interface, including the binary readouts
used to score evidence membership. Each candidate
is associated with a positive and negative decision
option, and the positive-minus-negative logit
margin defines its ranking score.

We use low-rank adaptation on the query, key,
value, and output projections of the model's eight
noncausal full-attention layers. The LoRA rank is
16 and its scaling parameter is 32. The existing
decision readout is also trainable, while the
remaining backbone parameters are frozen.
In total, approximately 3.8 million parameters
are updated.

Training interleaves two request types. Evidence
requests supervise multiple candidate decisions
using the masked, class-balanced loss in
Eq.~\ref{eq:joint_evidence_loss}. General-task
replay requests use the supervised and distillation
objective in Eq.~\ref{eq:replay_objective}.

Evidence requests constitute 60\% of sampled
training requests, while replay requests account
for the remaining 40\%. These proportions are
defined at the request level rather than the
individual readout level. Consequently, an
evidence request containing 128 candidate decisions
does not count as 128 separate training requests.

For general-task replay, the frozen pretrained
model supplies the teacher distribution over the
valid options of each task. Supervised labels and
teacher distributions receive equal objective
weights, with
$\lambda_{\mathrm{CE}}
=\lambda_{\mathrm{KD}}=0.5$.
The teacher is not used to supervise evidence
membership, since the objective is to improve
upon its pretrained evidence decisions rather
than reproduce them.

The final \textsc{CURATOR} checkpoint used in
our evaluations is obtained after 2,000 training
updates.


\subsection{Inference with Large Candidate Pools}
\label{app:curator_inference}

At inference time, the model receives a question
and a set of candidate evidence units. Each
candidate has a binary decision readout, and its
evidence score is computed using
Eq.~\ref{eq:evidence_margin}.

When the candidate set fits within the input
context window, all decisions are evaluated
jointly in one inference request. This avoids
repeatedly processing the full candidate context
for every individual decision.

Some retrieved pools exceed the model's context
window. In these cases, we divide the candidate
pool into smaller groups and perform joint scoring
within each group. In the full FEVEROUS evaluation,
we use groups of at most 200 candidates. The
resulting scores are combined into a single ranking
across the original candidate pool.

This distinction is important for interpreting
computational efficiency. Joint inference eliminates
repeated full-context evaluation within each group,
but a sufficiently large pool may still require
multiple inference requests. Therefore, large-pool
evaluation should not be interpreted as processing
every candidate in one model invocation.

Grouping also introduces a limitation: scores
produced in different groups do not share the
same complete candidate context. Their relative
calibration may differ from scores produced by
a hypothetical single inference over the entire
pool. In our matched FEVEROUS comparison, the
pretrained joint model and \textsc{CURATOR} use
the same grouping procedure.

Once all candidates have been scored, we sort
them in descending score order. Starting with
an empty selection, we admit a candidate if
its token cost fits within the remaining budget
$B$; otherwise, it is skipped. This continues
until all candidates have been considered.

The scoring and admission stages are separate.
The same computed scores can therefore be reused
across different token budgets without additional
model inference.

\subsection{Additional Training Ablations}
\label{app:curator_ablations}

We examine whether the improvement in evidence
discrimination comes from evidence-specific
fine-tuning alone or from training multiple
decisions jointly across different candidate
set sizes.

We compare pointwise evidence supervision,
joint supervision restricted to small candidate
sets, and joint supervision with variable $K$.
We also report the final \textsc{CURATOR}
checkpoint trained with general-task replay.

\begin{table}[t]
    \centering
    \small
    \setlength{\tabcolsep}{3pt}
    \caption{\textbf{Effect of evidence-training strategies.}
    Pooled evidence-membership AUROC on QAMPARI
    candidate sets of increasing size. The number
    of training updates is shown in parentheses.
    Higher is better.}
    \label{tab:curator_training_ablations}
    \begin{tabular}{@{}lrrr@{}}
        \hline
        Method & $K=8$ & $K=128$ & $K=256$ \\
        \hline
        Pretrained joint model
            & .7853 & .7076 & .6833 \\
        Pointwise (500)
            & .8578 & .7723 & .7560 \\
        Joint, $K\leq16$ (500)
            & .9057 & .9114 & .9104 \\
        Variable-$K$, no replay (500)
            & .9002 & .9322 & .9381 \\
        \textsc{CURATOR} (2000)
            & \textbf{.9104}
            & \textbf{.9450}
            & \textbf{.9541} \\
        \hline
    \end{tabular}
\end{table}

Table~\ref{tab:curator_training_ablations}
shows that pointwise evidence fine-tuning improves
discrimination relative to the pretrained model,
but performance still declines as the number of
jointly evaluated candidates increases.

Training on small joint candidate sets produces
a different behavior. Its AUROC remains relatively
stable as $K$ increases, suggesting that joint
evidence supervision is important for learning
to distinguish supporting evidence in shared
contexts.

Variable-sized joint supervision further improves
performance at larger candidate counts. Its AUROC
increases from 0.9002 at $K=8$ to 0.9381 at
$K=256$. Notably, $K=256$ exceeds the largest
candidate count used during training.

The final \textsc{CURATOR} checkpoint achieves
the highest discrimination among the evaluated
configurations. However, these runs differ in
training duration and other training choices.
The comparison therefore supports the importance
of joint and variable-sized supervision but
does not isolate the causal contribution of
each component under a fully matched setting.

\subsection{General Decision Capability Preservation}
\label{app:curator_capability}

Evidence-specific adaptation may alter the
pretrained model's behavior on general decision
tasks. We therefore evaluate whether the adapted
model retains performance across its original
decision formats.

The evaluation covers seven benchmarks:
BoolQ, RTE, MNLI, ARC-Easy, ARC-Challenge,
CommonsenseQA, and STS-B, comprising 19,631
examples in total. These benchmarks include
binary classification, multiple-choice
reasoning, and semantic similarity scoring.

Across the seven benchmarks, the unweighted
mean of their native evaluation scores is
0.8733 for the pretrained model and 0.8810
for \textsc{CURATOR}. This comparison shows
that evidence-specific adaptation does not
produce an overall decline in the measured
general-task performance.

These results are consistent with the
objective of retaining the model's general
decision capabilities during specialization.
However, they do not independently establish
that replay is responsible for this preservation.
A matched comparison with and without replay,
evaluated on the same general decision benchmarks,
would be needed to isolate its contribution.

\section{Detailed Ablation Studies}
\label{app:ablations}

We examine how pointwise supervision, joint training, candidate-set size, and general-task replay affect evidence discrimination. Table~\ref{tab:training_ablations} reports pooled AUROC on QAMPARI at different candidate counts.

\begin{table*}[t]
\centering
\small
\setlength{\tabcolsep}{8pt}
\renewcommand{\arraystretch}{1.1}
\caption{\textbf{Training ablations on QAMPARI.}
Pooled AUROC at different candidate counts.
Ablation variants use 500 updates; the selected CURATOR
checkpoint uses 2,000 updates. Higher is better.}
\label{tab:training_ablations}
\begin{tabular}{lccccc}
\hline
\textbf{Training Configuration}
& \textbf{Updates}
& $\boldsymbol{K=8}$
& $\boldsymbol{K=128}$
& $\boldsymbol{K=256}$
& $\boldsymbol{\Delta_{256-8}}$ \\
\hline
Original Joint Decider
& 0 & .7853 & .7076 & .6833 & $-.1020$ \\
Pointwise ($K=1$)
& 500 & .8578 & .7723 & .7560 & $-.1018$ \\
Joint ($K\leq16$)
& 500 & .9057 & .9114 & .9104 & $+.0047$ \\
Joint, variable $K$ (no replay)
& 500 & .9002 & .9322 & .9381 & $+.0379$ \\
\textbf{CURATOR (Ours)}
& 2000 & \textbf{.9104} & \textbf{.9450}
& \textbf{.9541} & $\mathbf{+.0437}$ \\
\hline
\end{tabular}
\end{table*}

\paragraph{Effect of joint supervision.}
Pointwise adaptation improves AUROC at small $K$ but exhibits nearly the same decline as the original model at larger candidate counts. Joint training with $K\leq16$ largely stabilizes discrimination, while variable-$K$ training further improves large-set performance. At matched 500-update training, the full variable-$K$ recipe improves mean per-question AUROC over small-$K$ training by 0.0148, averaged across $K=64$ and $K=128$ on QAMPARI.

\paragraph{Role of replay and experimental limitations.}
Removing general-task replay has only a small effect on evidence discrimination: the no-replay variant trails the matched 500-update full recipe by 0.0029 on the primary large-$K$ metric. However, its effect on preserving general capabilities is not established without a matched retention evaluation. Furthermore, pointwise adaptation uses QAMPARI-only training data, while joint variants use three datasets. The final checkpoint also receives more training updates. These comparisons therefore support the value of joint supervision without completely isolating each training component.

\section{Preservation of General Decision Capabilities}
\label{app:general_retention}

We evaluate whether adapting the pretrained decision model for evidence selection affects its original capabilities. We compare CURATOR with the original joint decider on seven general decision benchmarks covering inference, question answering, multiple-choice reasoning, and semantic similarity.

\begin{table}[t]
\centering
\footnotesize
\setlength{\tabcolsep}{3pt}
\renewcommand{\arraystretch}{1.1}
\caption{\textbf{Preservation of general decision capabilities.}
Performance before and after evidence adaptation
across seven benchmarks.}
\label{tab:general_retention}
\begin{tabular}{@{}lrrr@{}}
\hline
\textbf{Task}
& $\boldsymbol{n}$
& \textbf{Original}
& \textbf{CURATOR} \\
\hline
BoolQ          & 3,270 & .9153 & .9174 \\
RTE            & 277   & .9170 & .9278 \\
MNLI           & 9,815 & .9127 & .9151 \\
ARC-Easy       & 2,376 & .9903 & .9916 \\
ARC-Challenge  & 1,172 & .9531 & .9539 \\
CommonsenseQA  & 1,221 & .8608 & .8657 \\
STS-B          & 1,500 & .5640 & .5953 \\
\hline
\textbf{Macro-avg.}
& \textbf{19,631}
& \textbf{.8733}
& \textbf{.8810} \\
\hline
\end{tabular}
\end{table}

\paragraph{General-task performance.}
CURATOR achieves comparable performance to the original model across all seven benchmarks. Although its point estimates are slightly higher, none of the individual differences is statistically resolved. These results indicate no detectable degradation in the measured capabilities, rather than establishing equivalence or improved general-task performance.

\paragraph{Decision calibration.}
We additionally examine whether adaptation changes decision behavior on BoolQ. The predicted positive rate changes only from 0.6183 to 0.6205, while expected calibration error changes from 0.0136 to 0.0153. These results suggest that evidence adaptation preserves not only task performance but also broadly similar decision behavior on this benchmark.

\end{document}